\documentclass[a4paper]{JAC2003}    
\usepackage{graphicx}
\begin{document}

\title{BEAM-BEAM EFFECTS UNDER THE INFLUENCE OF EXTERNAL NOISE}
\author{K.~Ohmi, KEK, Oho, Tsukuba, 305-0801, Japan}

\maketitle
\begin{abstract}
Fast external noise, which gives fluctuation into the beam orbit, 
is discussed in connection with beam--beam effects.
Phase noise from crab cavities and
detection devices (position monitor) and kicker noise from
the bunch by bunch feedback system are the sources.
Beam-beam collisions with fast orbit fluctuations with turn by turn or 
multi-turn correlations, cause emittance growth and 
luminosity degradation. 
We discuss the tolerance of the noise amplitude for LHC and HL-LHC.
\end{abstract}

\section{INTRODUCTION}
Beam-beam effects under external noise are studied
with the weak-strong model in this paper.
The strong beam is regarded as a target with a Gaussian charge distribution.
In the model, an external noise is introduced into 
the transverse position
of the strong beam at the collision point. 

We first discuss 
an orbit (transverse position) shift of the strong beam 
given as:              
\begin{equation}
\Delta x_{i+1}=(1-1/\tau)\Delta x_{i}+\delta x\cdot \hat{r} \label{dxtype1}
\end{equation}
where $\Delta x_i$ is the orbit shift at the i$^{th}$ turn.
$\tau$, $\delta x$ and $\hat{r}$ are the damping times,
a constant characterizing the random fluctuation amplitude
and a Gaussian random number with unit standard deviation.
This is known as the Ornstein-Uhlenbeck process.
This type of noise is referred to as first type later.

All particles in the weak beam experience the fluctuation of the strong beam, 
thus  
a transverse collective motion is induced.
The collective motion results in an emittance growth 
due to filamentation caused by the nonlinear beam--beam force. 

The stable amplitude of the fluctuation of the strong beam is given by:
\begin{equation}
\Delta x^2=\langle \Delta x^2_{n\rightarrow \infty}\rangle
=\frac{\tau\delta x^2}{2}.\label{stabledx}
\end{equation}
The correlation function between i$^{th}$ and i$^{n-th}$ turns
is expressed by the damping time as:               
\begin{equation}
\langle \Delta x_\ell \Delta x_{\ell+n}\rangle=\Delta x^2 e^{-|n|/\tau}.
\end{equation}
The damping time is regarded as the correlation time of the fluctuation.
For white noise, which corresponds to $\tau=1$,
the correlation function is expressed as:
\begin{equation}
\langle \Delta x_\ell \Delta x_{\ell+n}\rangle=\Delta x^2 \delta_{n0}
\end{equation}
where $\delta_{n0}$ is the Kronecker delta.


The beam oscillates with the betatron frequency.
We consider a second type of noise as:            
\begin{equation}
\Delta x_{i+1}=(1-1/\tau)(\Delta x_{i}\cos \mu_o+\Delta p_{i}\sin \mu_o)
+\delta x \hat{r} \nonumber
\end{equation}
\begin{equation}
\Delta p_{i+1}=(1-1/\tau)(-\Delta x_{i}\sin \mu_o+\Delta p_{i}\cos \mu_o)
+\delta x \hat{r} \label{dxtype2}
\end{equation}
where $x$ and $p$ are the coordinate and canonical momentum normalized by
the beta function, so that $J=(x^2+p^2)/2$. $\mu_o=2\pi\nu_o$ is the betatron tune
multiplied by $2\pi$.
In collision the offset causes an emittance growth.
The stable dipole oscillation amplitude is expressed by the same 
equation as Eq.~(\ref{stabledx}).
The correlation function contains the betatron tune as:            
\begin{equation}
\langle \Delta x_\ell \Delta x_{\ell+n}\rangle
=\Delta x^2 e^{-|n|/\tau} \cos  n\mu_o .\label{corr2}
\end{equation}

We discuss the effect of noise for the cases of the LHC and High Luminosity-LHC.
The parameters are listed in Table 1.
The phenomena depend on the beam--beam parameter, the noise amplitude
normalized by the beam size and the Piwinski angle.  
\begin{table*}[htp]
\caption{Parameters for LHC (50~ns bunch spacing) and HL-LHC (25~ns bunch spacing).}
\label{LHCpara}
\begin{center}
\begin{tabular}{|l|ccc|}
\hline
  & LHC & HL-LHC(25ns) & HL-LHC(50ns)\\ \hline
Circumference (m) &  \multicolumn{3}{c|}{26 658} \\
Energy (TeV) &  \multicolumn{3}{c|}{7}\\
Tunes $Q_{x}, Q_{y}, Q_{s}$  & \multicolumn{3}{c|}{64.31,~59.32,~0.0019}   \\
Normalized Emittance ($\mu$m) & 2.0 & 2.5 & 3.0\\
$\beta^*$ (m) & 0.55 & 0.15 & 0.15\\
Bunch length (m) & \multicolumn{3}{c|}{0.0755} \\
Bunch population ($10^{11}$)
& $1.65$ & $2.2$ & $3.5$\\
Number of bunches & 1380 & 2808 & 1404 \\
Beam-beam parameter/IP  & 0.0034 & 0.005-0.011 & 0.005-0.014 \\
\hline
\end{tabular}
\end{center}
\end{table*}

\section{Emittance growth due to the external noise}
The emittance growth under an external noise and the non-linear force
of the beam--beam interaction is discussed
in \cite{Stupakov,TSen,AlexBBN}.
Previous work is reviewed in this section.

The beam--beam potential for a bunch population $N_p$ 
and the transverse size $\sigma_r$ is expressed as:
\begin{equation}
U(x)=\frac{N_pr_p}{\gamma_p}\int_0^{\infty} \frac{1-e^{-x^2/(2\sigma_r^2+q)}}
{2\sigma_r^2+q}dq
\end{equation}
where $r_p$ and $\gamma_p$ are the classical radius of the proton and the
relativistic factor of the (weak) beam, respectively.
The potential is expanded as a Fourier series:               
\begin{equation}
U(x)=\frac{N_pr_P}{\gamma_p}\sum^{\infty}_{k=0} U_k(a) \cos 2k\psi 
\end{equation}
where
\begin{equation}
U_k(a)=\int_0^{a} \left[\delta_{0k}-(2-\delta_{0k})(-1)^ke^{-w}I_k(w)\right]
\frac{dw}{w},
\end{equation}
and $a=\beta^* J/2\sigma_r^2=J/2\varepsilon$.
The change of $J$ per revolution is given by the derivative of the
beam--beam potential with respect to $\psi$ as:          
\begin{equation}
\Delta J=-\frac{\partial U}{\partial \psi}
=\frac{Nr_p}{\gamma}\sum_{k=0}^\infty 2k U_k \sin 2k\psi.
\end{equation}
This change, which indicates a stable sinusoidal modulation of 
the betatron amplitude, does not induce emittance growth.

We consider the case in which
the strong beam has a small offset ($\Delta x$). 
The beam--beam potential with the offset is
expanded for $\Delta x$:
\begin{equation}
U(x+\Delta x)=U(x)+U'(x)\Delta x.
\end{equation}
Here $\Delta x$ is a random variable fluctuating described
by Eq.~(\ref{dxtype1}) or (\ref{dxtype2}).

The potential with the offset is expanded as a Fourier series:
\begin{eqnarray}
\lefteqn{U'(J,\psi)=\frac{\partial U}{\partial J}\frac{\partial J}{\partial x}+\frac{\partial U}{\partial \psi}\frac{\partial \psi}{\partial x}}\\
&& =\frac{N_pr_p}{2\gamma\sigma_r}\sum_{k=0}^{\infty}G_k(a)\cos(2k+1)\psi.
\end{eqnarray}
The Fourier coefficients as a function of $a$ are expressed as:
\begin{equation}
G_k(a)=\sqrt{a}\left[U'_{k+1}+U'_k\right]
+\frac{1}{\sqrt{a}}\left[(k+1)U_{k+1}-kU_k\right].
\end{equation}
where $U'_k$ is the derivative with respect to $a$.

The diffusion of $J^2$ after N revolutions is given by:
\begin{equation}
\langle\Delta J^2(N)\rangle=\sum_{\ell=1}^N\sum_{n=-\ell+1}^{N-\ell}
\frac{\partial U'(\ell)}{\partial\psi}
\frac{\partial U'(\ell+n)}{\partial\psi}
\langle \Delta x_\ell\Delta x_{\ell+n}\rangle
\end{equation}

For turn-by-turn white noise, the correlation function is replaced by
the Kronecker delta, $\delta_{n0}$. 
The diffusion of $J$ is expressed by:
\begin{equation}
\langle\Delta J^2\rangle=\frac{\langle\Delta J^2(N)\rangle}{N}
\approx \frac{N_p^2r_p^2}{8\gamma^2\sigma_r^2}\sum_{k=0}^{\infty}
(2k+1)^2G_k(a)^2 .
\end{equation}

The diffusion of $J$ per revolution is given for the fluctuation 
in Eq.~(\ref{dxtype1}) by:
\begin{eqnarray}
\lefteqn{\langle\Delta J^2\rangle
\approx\frac{N_p^2r_p^2}{8\gamma^2\sigma_r^2}}\nonumber \\&&
\sum_{n=-\infty}^{\infty}\sum_{k=0}^{\infty}(2k+1)^2G_k^2
\cos[(2k+1)n\mu_o] e^{-|n|/\tau} \nonumber \\
&&\approx \frac{N_p^2r_p^2}{8\gamma^2\sigma_r^2}\sum_{k=0}^{\infty}
\frac{(2k+1)^2G_k(a)^2 \sinh 1/\tau}{\cosh 1/\tau-\cos(2k+1)\mu_o}.
\label{T1formula}
\end{eqnarray}

The diffusion of $J$ for the second type of noise (Eq.~(\ref{dxtype2})) 
is given using the correlation of Eq.~(\ref{corr2}):                 
\begin{eqnarray}
\lefteqn{\langle\Delta J^2\rangle
\approx \frac{N_p^2r_p^2}{16\gamma^2\sigma_r^2}\sum_{k=0}^{\infty}
(2k+1)^2G_k(a)^2 \sinh 1/\tau} \nonumber \\
&&
\left[\frac{1}{\cosh 1/\tau-\cos (2k\mu-\delta\mu)}+\right. \label{T2formula} \\
&&\left.\frac{1}{\cosh 1/\tau-\cos (2(k+1)\mu+\delta\mu)}\right]\nonumber
\end{eqnarray}
where $\delta \mu$ is the tune difference between the weak and strong beam 
oscillations ($\delta\mu=\mu-\mu_o$).

\begin{figure}[htp]
\centering
\scalebox{1}{\includegraphics{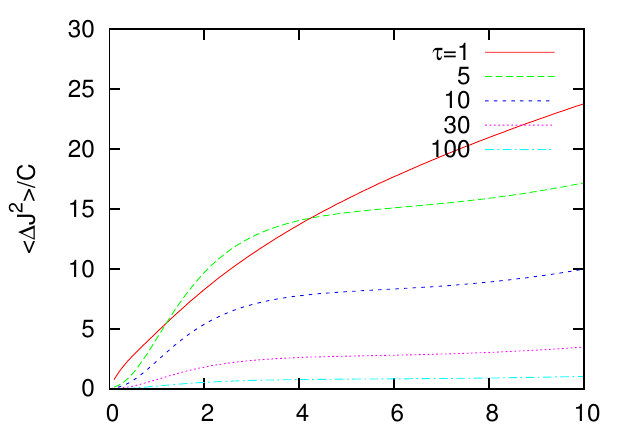}}
\caption{
Diffusion rate given by Eq.~(\ref{T1formula}). The rate is normalized by 
$C=(N_pr_p\Delta x/\gamma\sigma_r)^2/8$.
}
\label{DiffRate_TSen}
\end{figure}

Figure \ref{DiffRate_TSen} shows the diffusion rate of $J$ as a function of
$J$.
The diffusion rate is proportional to
the square of the fluctuation amplitude $\Delta x$
and the square of the beam--beam parameter $\propto N_p$.
The rate is normalized by the factor  
$C=(N_pr_p\Delta x/\gamma\sigma_r)^2/8$ in the figure.

The emittance growth is evaluated from the diffusion rate, when
the rate $\langle\Delta J^2\rangle$ is proportional to $J$.
\begin{equation}
\frac{\Delta\varepsilon}{\varepsilon}=
\frac{\langle\Delta J^2\rangle}{2\varepsilon J}=
\frac{1}{4\varepsilon^2}\frac{d\langle\Delta J^2\rangle}{da}.
\end{equation}

Figure \ref{DiffRate_TSen} shows that the rate is proportional to $J$
for small $J/2\varepsilon<2$. 
The slope of $\langle\Delta J^2\rangle$ for turn-by-turn noise
($\tau~=~1$) is:
\begin{equation}
\frac{\langle\Delta J^2\rangle}{a}=\frac{N_p^2r_p^2}{8\gamma^2}
\frac{\Delta x^2}{\sigma_r^2}
\times 4.4.
\end{equation}

The luminosity degradation rate per collision is estimated by the emittance
growth rate as:            
\begin{equation}
\Delta L/L=\left(\xi \frac{\Delta x}{\sigma_r}\right)^2\times 21.7.
\label{SimpleFormula0}
\end{equation}
For two IPs, the formula is corrected by a factor two, i.e. $21.7\rightarrow 10.8$
and $\xi\rightarrow \xi_{tot}$,
The tolerance for the noise amplitude is given for
a luminosity life time $\Delta L/L=10^{-9}$: 
\begin{equation}
\xi_{tot}\frac{\Delta x}{\sigma_r}=9.8\times 10^{-6}.\label{SimpleFormula}
\end{equation}

\begin{figure}[htp]
\centering
\scalebox{1}{\includegraphics{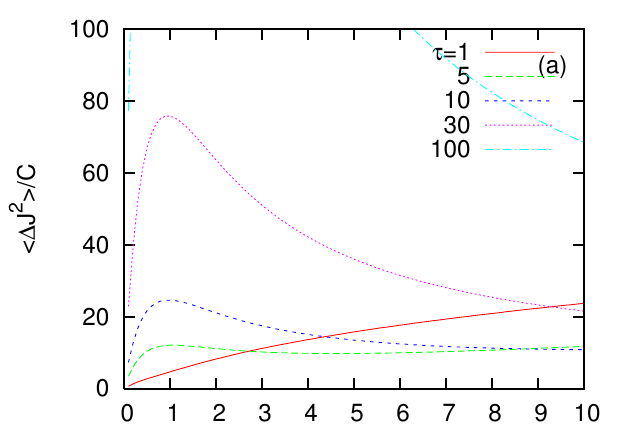}}
\scalebox{1}{\includegraphics{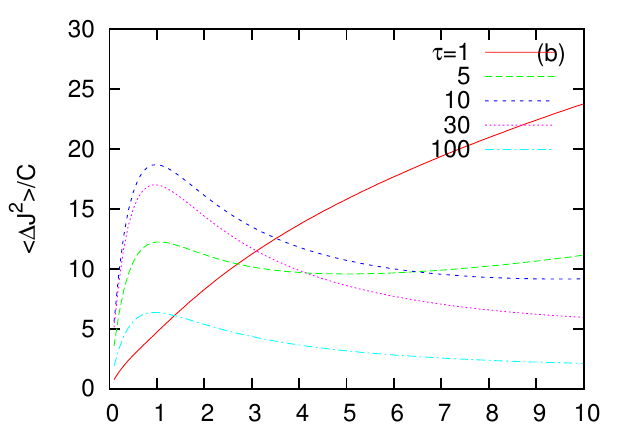}}
\caption{
Diffusion rate given by Eq.~(\ref{T2formula}). The rate is normalized by 
$C=(N_pr_p\Delta x/\gamma\sigma_r)^2/8$.
}
\label{DiffRate_Type2}
\end{figure}

We now discuss the second type of noise given by Eq.~(\ref{dxtype2}).
Figure \ref{DiffRate_Type2} shows the diffusion rates.
Figures~\ref{DiffRate_Type2}~(a) and (b) are given  for
the beam-orbit oscillation with the same tune ($\delta\mu=0$) 
and a difference of 
$\delta\mu=\xi=0.01$, respectively.
A strong enhancement  of the diffusion is seen
at small amplitudes at a large correlation time in shown in Fig.~\ref{DiffRate_Type2}~(a).
This behavior mainly comes from a contribution at $k=0$. 
\begin{equation}
\langle\Delta J^2\rangle
\approx \frac{N_p^2r_p^2}{16\gamma^2\sigma_r^2}G_0(a)^2 \tau 
\end{equation}
The strong beam modulation with the same tune  gives an external
force oscillation to the weak beam particles.
For colliding beams, the assumption, that beam-orbit oscillations
have the same tune, is not obvious.
The diffusion rate for $\delta\mu=\xi$ in Fig.~\ref{DiffRate_Type2}~(b) may be better  
to represent the beam--beam system.
The diffusion rate, which is saturated at $J/2\varepsilon=1$, 
is similar as that of $\tau=1$ on the whole.
Therefore we study the diffusion rate for $\tau=1$ in simulations.

It may be better that the noise effects are studied 
in the framework of a strong-strong model, especially 
for the second type of noise. 
The noise induces either coherent $\sigma$ or $\pi$ modes
or a continuous frequency spectrum. 
The $\sigma$ mode does not contribute
the emittance growth.
Emittance growth based on the strong-strong model had been discussed
in \cite{AlexBBN}.
The author discussed that 
18\% of the dipole motion induced by offset collision
into the  mode with continuous frequency spectrum.
The emittance increases by smearing the dipole motion.
The growth rate is expressed by:
\begin{equation}
\frac{\delta\varepsilon}{\varepsilon}\approx 
\frac{K}{\left(1+\frac{1}{2\pi\tau|\xi|}\right)^2}\frac{\delta x^2}{\sigma_x^2}
=\frac{2K}{\tau\left(1+\frac{1}{2\pi\tau|\xi|}\right)^2}
\frac{\Delta x^2}{\sigma_x^2}
\label{emigrA}
\end{equation}
where $K=0.089$ is a form factor for the emittance change
induced by a dipole amplitude, 
and the damping rate $1/\tau$ of the coherent motion. 
The emittance growth rate is independent of the beam--beam tune shift,
when $1/\tau\ll 2\pi|\xi|$, while the rate is proportional to the square of
the beam--beam tune shift, when $1/\tau\gg 2\pi|\xi|$.

\begin{figure}[htp]
\centering
\scalebox{1}{\includegraphics{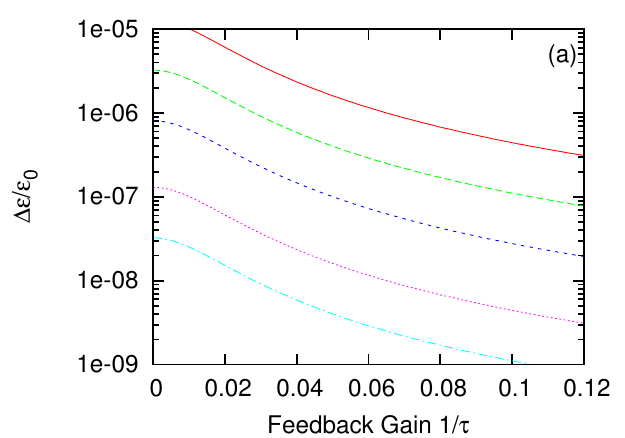}}
\scalebox{1}{\includegraphics{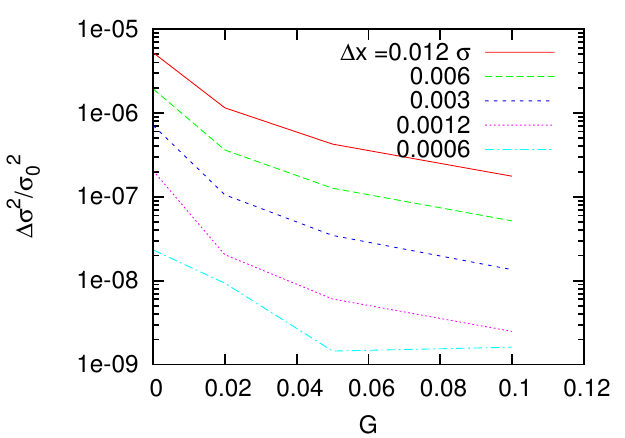}}
\caption{
Emittance growth given by Eq.~(\ref{emigrA})
and by a strong-strong beam--beam simulation \cite{ohmiBBN:PAC07}.
}
\label{emitgrSS}
\end{figure}

Figure \ref{emitgrSS} shows the emittance growth given by Eq.~(\ref{emigrA})
and by a strong-strong beam--beam simulation \cite{ohmiBBN:PAC07},
where the beam--beam tune shift is $\xi=0.0034/IP$.
The results agree fairly well.
The strong-strong simulation suffers from numerical noise related to the statistics
of macro-particles. 
One million macro-particles are used in the simulation, thus 
0.1\% of the offset noise is induced by the statistics.

\section{Simulation of external noise}
\subsection{Study based on LHC}
The analytical theory is based on the near solvable system
far from resonances. 
There is no such limitation in beam--beam simulations,
while simulations take considerable computing time 
to evaluate a slow emittance growth.
Simulations considering external noise are straightforward:
a modulation is applied to the strong beam with 
Eq.~(\ref{dxtype1}) or (\ref{dxtype2}).
Effects of resonances, longitudinal motion and a crossing angle
are taken into account in simulations.

We only discuss weak-strong simulations taking into account  
external noise. The weak beam is represented by 131072 macro-particles.
The particles are tracked one million turns interacting a strong beam located
at two interaction points. 
The luminosity is calculated turn-by-turn, and averaged every 100 turns.
Luminosity degradation is evaluated by fitting its evolution.
 
\begin{figure}[htp]
\centering
\scalebox{1}{\includegraphics{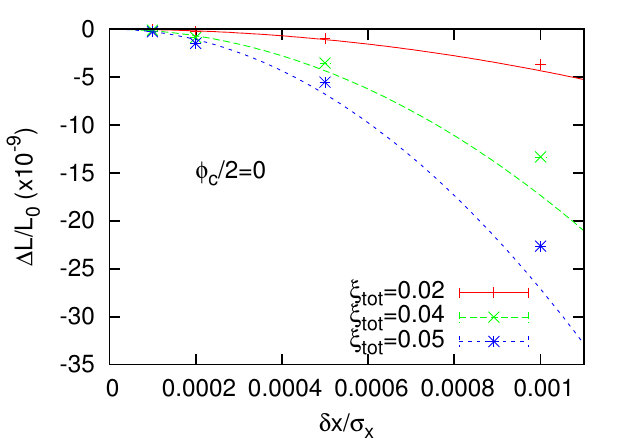}}
\caption{
Diffusion rate given by a weak-strong simulation using Eq.~(\ref{SimpleFormula}). 
}
\label{Ldamp_dx}
\end{figure}

Figure \ref{Ldamp_dx} shows the
luminosity degradation for collisions without a crossing angle.
The degradation is plotted as a function of the fluctuation amplitude
for three total beam--beam parameters, $\xi_{tot}=0.02$, 0.04 and 0.05. 
Three lines given by the analytical formula
Eq.~(\ref{SimpleFormula}) are shown in the figure.
The simulation results agree with the formula fairly well.

\begin{figure}[htp]
\centering
\scalebox{1}{\includegraphics{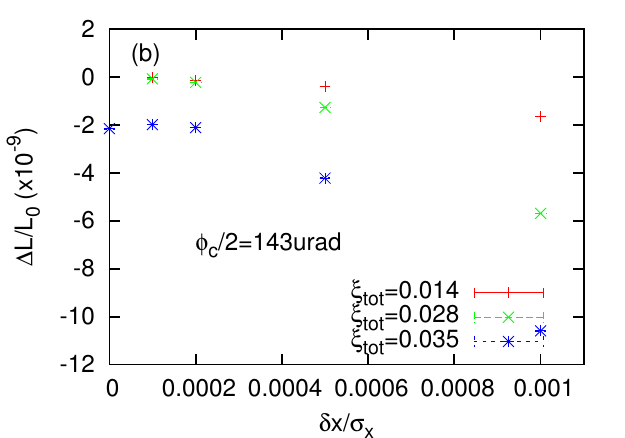}}
\caption{
Diffusion rate for crossing collision
given by weak-strong simulation. 
}
\label{LdampCrs_dx}
\end{figure}

The luminosity degradation for collision with a crossing angle 
($\phi_c=290$~$\mu$rad) 
is shown in Fig.~\ref{LdampCrs_dx}. 
The Piwinski angle is $\phi_c\sigma_z/2\sigma_r=0.89$.

\begin{figure}[htp]
\centering
\scalebox{1}{\includegraphics{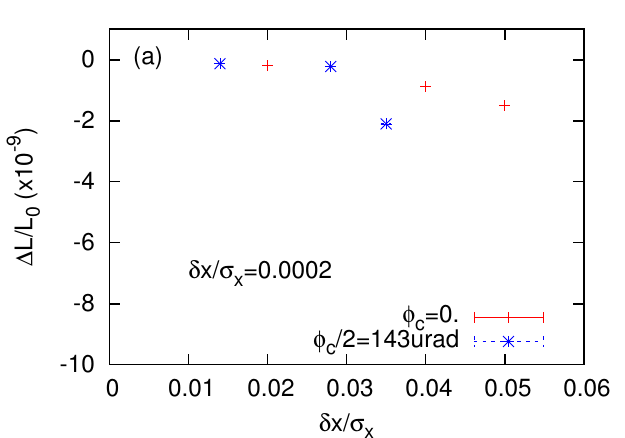}}
\scalebox{1}{\includegraphics{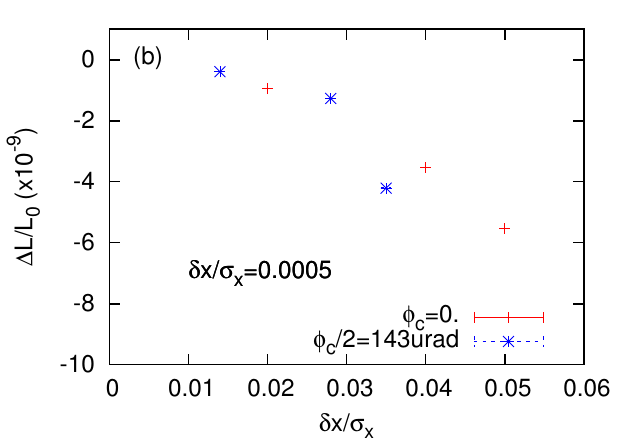}}
\caption{
Luminosity degradation as a function of the beam--beam parameter 
with offset noise.
}
\label{lumtnoise_xi}
\end{figure}

Figure \ref{lumtnoise_xi} shows luminosity degradation as a function of 
the beam--beam parameter under offset noise. 
The tune shift is reduced to 70\% for the crossing collision. 
The luminosity degradation for noise is independent of the crossing angle. 
At high beam--beam parameters $>0.05$, 
the luminosity degradation due to the crossing angle is dominant.

There was no qualitative change for collision without a crossing angle.
For $\xi_{tot}=0.035$, a degradation due to the crossing angle is seen,
but a significant cross-talk is not observed.
The degradation of the luminosity due to the fluctuation
depends on $\xi_{tot}$, but hardly on the presence of the crossing angle.

\subsection{High Luminosity LHC (HL-LHC)}
For the HL-LHC, a higher luminosity is the target and obtained by
increasing the bunch population and squeezing to smaller beta function, 
while the pile up of collision events sets an upper limit for the luminosity at
$L/coll=2.6\times 10^{31}$~cm$^{-2}$s$^{-1}$.
The luminosity at $\beta=0.15$~m is expected to be $L/coll=8.6$ or 
$18\times 10^{31}$~cm$^{-2}$s$^{-1}$
for the bunch population of 
2.2 or 3.5$\times 10^{11}$, respectively.
Therefore luminosity levelling keeping the luminosity constant at
$L/coll=2.6\times 10^{31}$~cm$^{-2}$s$^{-1}$ is proposed.
The levelling can be done by controlling the crab cavity voltage or the 
beta function at the interaction point (IP).
Leveling with the beta function, the total beam--beam parameter (2IP) 
is $0.011\times 2=0.022$ (25ns) or $0.014\times 2=0.028$ (50ns) 
at the early stage of the collision, where
the beta function is 0.49 m or 1.02 m.
The results given in the previous subsection are applied for the
parameters: 
\begin{equation}
\frac{\Delta x}{\sigma_r}=4.5\times 10^{-4}~\mbox{ or }~ 3.5\times 10^{-4}.
\label{ToreAmp}
\end{equation}
for 25 ns and 50 ns, respectively.

Using a levelling with crab cavities, 
the crab voltage increases to keep the luminosity constant while
the beam current decreases. 
At the early stage of collision, the crab voltage is low and two beams
collide with a large Piwinski angle, 
where $\phi_c\sigma_z/2\sigma_r=3.14$ or 2.87 for 25 ns or 50 ns, 
respectively.
We study the effects of noise for collisions
with a large Piwinski angle.

\begin{figure}[htp]
\centering
\scalebox{1}{\includegraphics{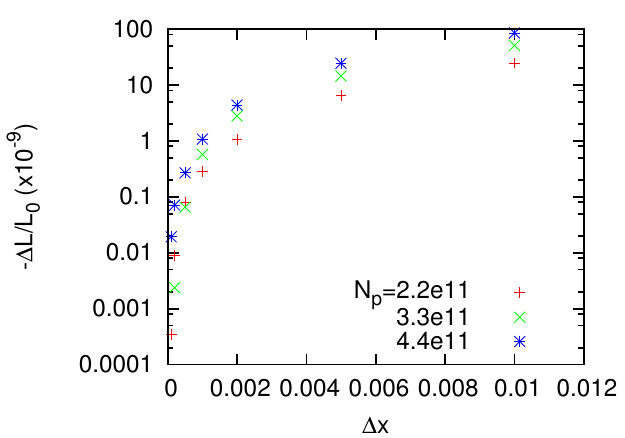}}
\caption{
Luminosity degradation as a function of noise amplitude.
}
\label{LumNoiseNp}
\end{figure}

Figure~\ref{LumNoiseNp} shows the luminosity degradation rate as a function
of the offset amplitude. 
The simulation is performed for two IPs. The tune shift is
0.0015 or 0.0050 in the crossing or orthogonal plane
for the design bunch population of $N_p=2.2\times 10^{11}$ (25ns).
The tune shift is 0.0065 in both planes, due to the combination of the 
horizontal and vertical crossing.
The fluctuation
amplitude 0.2\% is a tolerable limit for $\Delta L/L_0=10^{-9}$
as shown in the figure.
The simple formula Eq.~(\ref{SimpleFormula}) is satisfied for the HL-LHC, 
$0.0065\times 0.002=1.3\times 10^{-5}$, with 30\% difference from the formula.

\begin{figure}[htp]
\centering
\scalebox{1}{\includegraphics{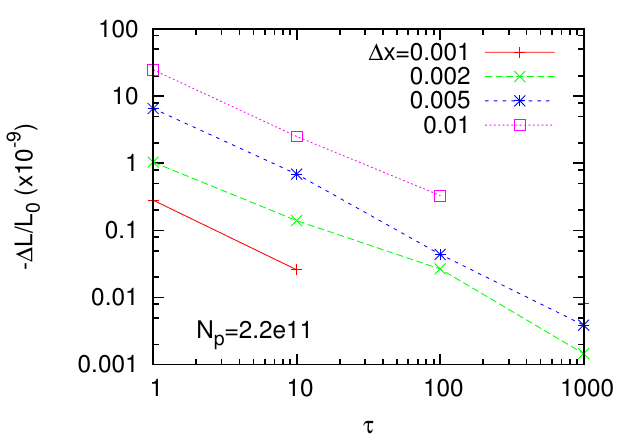}}
\caption{Luminosity degradation as a function of the correlation time.
}
\label{LumNoiseTcor}
\end{figure}

Figure \ref{LumNoiseTcor} shows the luminosity degradation as a
function of the correlation time. 
The luminosity degradation, which scales as $1/\tau$, is
consistent with Eq.~(\ref{T1formula}).

The effect of noise 
on its amplitude and the dependence on the correlation time
is similar for collisions with a large Piwinski angle 
and those for ordinary collisions without a crossing angle.
The luminosity degradation depends on the beam--beam parameter and the noise amplitude,
but with little dependence on the Piwinski angle.

\subsection{Tolerance for crab cavity phase noise in HL-LHC}
Crab cavities are used to compensate the crossing angle 
($\phi_c=590$~$\mu$rad) at IP.
The relation of the phase noise and collision offset is given by
\begin{equation}
\Delta \varphi_{cc}=\frac{\omega_{cc}}{c \phi_c/2}\Delta x,
\end{equation}
where $\Delta\varphi_{cc}$ and $\omega_{cc}$ are the phase fluctuation
and frequency of the crab cavity.

Using beta function levelling the beam--beam parameter is very high,
$\xi_{tot}=0.022$ or 0.028
for 25 ns or 50 ns, respectively.
The tolerance of the noise amplitude is given by Eq.~(\ref{ToreAmp}).
The corresponding phase error is $\Delta\varphi=1.6\times 10^{-4}$ or
$2.3\times 10^{-4}$ rad.

For the crab cavity levelling, the beam--beam parameter is $\xi_{tot}=0.0065$.
The tolerance of the noise amplitude
is $\Delta x/\sigma_r=0.002$ and the corresponding phase error is 
$\Delta\varphi=4\times 10^{-3}$, where the crab angle is 10\% of 
the crossing angle $\phi_c=59$~$\mu$rad 
($L/coll=2.7\times 10^{31}$~cm$^{-2}$s$^{-1}$).

The crab cavity noise was measured at KEKB, 
$1.7\times~10^{-4}$~rad for frequencies above 1~kHz ($\tau<10$).
The value is critical for beta function levelling, 
because of the large beam--beam parameter.
Using four crab cavities, the noise tolerance is twice as large,      
while for the crab voltage levelling, the measured phase error is tolerable.

\subsection{Incoherent noise due to intra-beam scattering}
Emittance growth times due to intra-beam scattering (IBS) are 105 h and 63 h 
for the horizontal and longitudinal planes, respectively, in the nominal LHC [13]. 
The transverse emittance and bunch population in the nominal are 
$5.0\times 10^{-10}$ and $1.15\times 10^{11}$, respectively. 
The horizontal IBS growth rate is approximately proportional 
to the particle density in the six dimensional phase space. 
The growth time is 40 h for $\xi_{tot}$~=~0.02 in this paper 
($\varepsilon=2.7\times 10^{-10}$ and $Np=1.63\times 10^{11}$). 
The fluctuation is $\delta x/\sigma_x=5.5\times 10^{-5}$ for 
$\xi_{tot}$~=~0.05. 
The luminosity degradation is determined by geometrical emittance 
growth $\delta L/L_0= \delta x^2/\sigma_x^2$ for incoherent noise.

\section{Coherent beam--beam effects under external noise}
Effects of external noise in crab cavity were performed in KEKB
during 2008 and 2009 \cite{nlcohbb}. 
Sinusoidal noise is applied to the crab cavity RF system.
Near the $\sigma$ mode tune, a strong luminosity drop of 80~\% was seen
when suddenly exceeding a threshold excitation amplitude.
A smaller luminosity drop ($L=0.9 L_0$) was seen near $\pi$ mode frequency.
Strong-strong simulations reproduced these luminosity drops.
A systematic study using the strong-strong simulation showed that
these characteristic phenomena for coherent 
nonlinear beam--beam interactions. A similar phenomenon was observed
in Ref.\cite{IeiriHirata}.
The detailed analysis is published in Ref.\cite{nlcohbb}. 

\section{CONCLUSIONS}
Fast noise of the collision offset degrades the luminosity performance
in hadron colliders.
The luminosity degradation depends on the product of the noise amplitude 
and the beam--beam parameter as shown in Eq.~(\ref{SimpleFormula}),
with little dependence on the crossing angle.
A tolerance for the crab cavity phase error was obtained for the HL-LHC.

The crab cavity noise was measured at KEKB, 
$1.7\times 10^{-4}$ rad above 1~kHz ($\tau<10$).
The value is critical for beta function levelling, 
because of the high beam--beam parameter.
For the crab voltage levelling, the measured phase error is tolerable
because of the small beam--beam tune shift.

Further studies related to beam--beam modes should be done using 
strong-strong models.

\section{Acknowledgement}
The author would like to thank 
Drs. Y. Alexahin, R. Calaga, W. Herr, S. Paret, T. Pieloni, 
J. Qiang, R. Tomas  and F. Zimmermann for fruitful discussions.


\end{document}